\documentclass[aps,pra,reprint, amsmath, amssymb,superscriptaddress,nofootinbib]{revtex4-1}

\usepackage{bm}
\usepackage[retainorgcmds]{IEEEtrantools}
\usepackage{graphicx}
\usepackage{mathrsfs}
\usepackage{amsmath}
\usepackage{amssymb}
\usepackage{color}
\usepackage{amsfonts}
\usepackage{times,txfonts}
\usepackage{nicefrac}
\usepackage[colorlinks=true,linkcolor=blue,urlcolor=blue,citecolor=blue,pdfusetitle]{hyperref}
\usepackage{amsmath}

\newcommand{\tr}{\text{tr}}

\begin{document}

\title{Moment generating function bound from detailed fluctuation theorem}
\date{\today}
\author{Domingos S. P. Salazar}
\affiliation{Unidade de Educa\c c\~ao a Dist\^ancia e Tecnologia,
Universidade Federal Rural de Pernambuco,
52171-900 Recife, Pernambuco, Brazil}

\begin{abstract}
A famous consequence of the detailed fluctuation theorem (FT), $p(\Sigma)/p(-\Sigma)=\exp{(\Sigma)}$, is the integral FT $\langle \exp(-\Sigma)\rangle =1$ for a random variable $\Sigma$ and a distribution $p(\Sigma)$. When $\Sigma$ represents the entropy production in thermodynamics, the main outcome of the integral FT is the second law, $\langle \Sigma \rangle \geq 0$. However, a full description of the fluctuations of $\Sigma$ might require knowledge of the moment generating function (MGF), $G(\alpha):=\langle \exp(\alpha \Sigma) \rangle$. In the context of the detailed FT, we show the  MGF is lower bounded in the form $G(\alpha)\geq B(\alpha,\langle\Sigma\rangle)$ for a given mean $\langle\Sigma\rangle$.  As applications, we verify that the bound is satisfied for the entropy produced in the heat exchange problem between two reservoirs mediated by a weakly coupled bosonic mode and a qubit swap engine.
\end{abstract}
\maketitle{}

%\section{Introduction}

% {\bf \emph{Introduction -}}
{\bf \emph{Introduction -}} 
The second law of thermodynamics states that the entropy production is nonnegative. Although the random nature of $\Sigma$ is practically negligible in large systems, it dominates the physics of the entropy production at small scale due to thermal and quantum fluctuations \cite{RevModPhys.93.035008,Seifert2012Review,Campisi2011,Esposito2009,Jarzynskia2008,Jarzynski1997,Ciliberto2013,Crooks1998,Gallavotti1995,Evans1993,Hanggi2015,Batalhao2014,Pal2019}. In this context, the second law is stated as an average, $\langle \Sigma \rangle \geq 0$. 

The detailed fluctuation theorem (DFT) is a stronger statement, which defines a random variable (typically the entropy production) in terms of the ratio of probabilities,
\begin{equation}
\label{DFT}
  \Sigma(\Gamma):=\ln P(\Gamma)/P(\Gamma^\dagger),
\end{equation}
where $\Gamma$ is some process and $\Gamma^\dagger$ is a conjugate (involution), such that $\Gamma^{\dagger\dagger}=\Gamma$. As a consequence of (\ref{DFT}), for instance, we have the integral FT, $\langle \exp(-\Gamma)\rangle =1$ and the second law, $\langle \Sigma \rangle \geq 0$, from Jensen's inequality. The definition (\ref{DFT}) is used to define the trajectory level entropy production \cite{Seifert2005} in stochastic thermodynamics, and a form of (\ref{DFT}) is also called the strong detailed fluctuation theorem \cite{Seifert2012Review}, Evan-Searles fluctuation theorem \cite{Evans2002}, Gallavotti-Cohen relation \cite{Gallavotti1995} and it appears in the heat exchange problem \cite{Jarzynski2004a}. The structure of (\ref{DFT}) might also be used to define generalizations of entropy production that contains information terms \cite{Potts2019} and in quantum systems beyond the two point measurement scheme \cite{Micadei2019}. For that reason, we treat (\ref{DFT}) as a definition of a general random variable.

Beyond the second law, the DFT (\ref{DFT}) has known consequences for the statistics of $\Sigma$. For instance, a form of Thermodynamic Uncertainty Relation (TUR) can also be derived from it \cite{Merhav2010,Timpanaro2019B,Hasegawa2019,Salazar2022d}, which can be understood as a lower bound for the variance in terms of the mean of $\Sigma$. Another example is a lower bound for apparent violations of the second law \cite{Salazar2021b}, defined as the cases for which $\Sigma < 0$.

The randomness of $\Sigma$ is encoded in the moment generating function (MGF),
\begin{equation}
\label{mgf}
G(\alpha):=\langle e^{\alpha \Sigma} \rangle = \sum_{\Gamma} e^{\alpha \Sigma (\Gamma)}P(\Gamma),
\end{equation}
where the sum above might be replaced by an integral and $\alpha$ is a real number. Derivatives of (\ref{mgf}) at $\alpha=0$ have information of statistical moments. For instance, the first derivative is related to the second law, $G'(0)=\langle \Sigma \rangle\geq 0$. In general, we have $G^{(n)}(0)=\langle \Sigma^n\rangle$ for higher order moments. The MGF (\ref{mgf}) is nonnegative, $G(\alpha)\geq 0$, but a stronger bound comes from Jensen's inequality, where $G(\alpha)\geq \exp(\alpha \overline{\Sigma})\geq 0$.

In this letter, we propose the following question: what is the impact of the DFT (\ref{DFT}) in the MGF (\ref{mgf})? The first immediate known consequence of (\ref{DFT}) is the known Evan-Searles or Gallavotti-Cohen symmetry property for (\ref{mgf}),
\begin{equation}
\label{symmetry}
G(\alpha)=G(-1-\alpha),
\end{equation}
as a direct application  of the definition (\ref{DFT}). Moreover, in terms of the MGF (\ref{mgf}), the IFT for $\Sigma(\Gamma)$ is obtained for $\alpha=-1$ using (\ref{symmetry}),
\begin{equation}
\label{ift}
\langle e^{-\Sigma} \rangle=G(-1)= G(0)= 1,
\end{equation}
and the second law follows from it. Our main result is to show that the DFT (\ref{DFT}) implies a tight lower bound for the MGF (\ref{mgf}) for a given mean $\overline{\Sigma}$,
\begin{equation}
\label{main}
G(\alpha) \geq \frac{\cosh[(\alpha+1/2)g(\overline{\Sigma})]}{\cosh[(1/2)g(\overline{\Sigma})]},
\end{equation}
for $\alpha \in \mathbb{R}$,
where $g(x)$ is the inverse function of $h(x):=x\tanh(x/2)$, defined for $x\geq0$ and $\overline{\Sigma}:=\langle \Sigma \rangle=\sum_\Gamma \Sigma(\Gamma)P(\Gamma)$. We show that the distribution that saturates (\ref{main}) is the distribution that saturates the TUR \cite{Timpanaro2019B}. This bound explores the DFT (\ref{DFT}) directly to improve on the bound $\exp(\alpha \overline{\Sigma})$.

The letter is organized as follows. First, we present a general formalism for (\ref{DFT}) in terms of involutions to derive (\ref{main}). Then, we discuss how the result can be framed using information theory in terms of Rényi relative entropy. As applications, we show that the bound is satisfied in three different cases: a Gaussian distribution, and the heat exchanged between two reservoirs mediated by a bosonic mode in weak coupling and a qubit swap engine.

%\section{Formalism}
{\bf \emph{Formalism -}}
Let $\Gamma \in S$ be an element of a set $S$ and $m:S\rightarrow S$ is any involution $m(m(\Sigma))=\Sigma$, where we define the notation
$\Gamma^\dagger:=m(\Sigma)$. Let $P:\Gamma \rightarrow [0,1]$ be a probability function such that $P(\Gamma)=0 \rightarrow P(\Gamma^\dagger)=0$ (absolute continuity). For instance, one could choose $S=\mathbb{R}$ and $m(x)=-x$. More generally, in 
stochastic thermodynamics, $\Gamma=(x_1,...,x_n) \in \mathbb{R}^n$ is a sequence and $\Gamma^\dagger = (x_n,...,x_1)$ is the inverse. 

Define the function $\Sigma:S\rightarrow \mathbb{R}$ as $\Sigma(\Gamma):=\ln(P(\Gamma)/P(\Gamma^\dagger))$ for $P(\Gamma)>0$ as in (\ref{DFT}) and $\Sigma(\Gamma):=0$, for $P(\Gamma)=P(\Gamma^\dagger)=0$. Usually, this expression for $\Sigma(\Gamma)$ appears in nonequilibium thermodynamics as the stochastic entropy production or some generalization. The moment generating function of $\Sigma(\Gamma)$ is given by (\ref{mgf})
\begin{equation}
    \label{mgf2}
    G(\alpha)= \sum_{\Gamma} e^{\alpha \Sigma(\Gamma)}P(\Gamma)=\sum_{\Gamma} \int_{-\infty}^{\infty}  e^{\alpha \sigma} \delta(\Sigma(\Gamma)-\sigma)P(\Gamma)d\sigma,
\end{equation}
where we used $f(x)=\int  f(y)\delta(y-x)dy$ and $\delta(x)$ is the Dirac's delta function. Now define the pdf
\begin{equation}
    \label{pdf}
    p(\sigma):=\sum_{\Gamma} \delta(\Sigma(\Gamma)-\sigma)P(\Gamma),
\end{equation}
such that the mgf (\ref{mgf2}) is given as
\begin{equation}
    \label{mgf2}
    G(\alpha)= \int_{-\infty}^{\infty} e^{\alpha \sigma}p(\sigma)d\sigma=\langle e^{\alpha \sigma}\rangle,
\end{equation}
where $\langle \rangle$ are now understood as averages in $p(\sigma)$. Similarly, we have $\langle \sigma\rangle = \langle \Sigma \rangle$. From the definition of $\Sigma(\Gamma)$, we have the property for $p(\sigma)$
\begin{equation}
p(\sigma)=\sum_{\Gamma} \delta(\Sigma(\Gamma)-\sigma)e^{\Sigma(\Gamma)}P(\Gamma^\dagger)=e^{\sigma}\sum_{\Gamma}\delta(\Sigma(\Gamma)-\sigma)P(\Gamma^\dagger),
\end{equation}
and using $\Sigma(\Gamma)=-\Sigma(\Gamma^\dagger)$, definition (\ref{pdf}) and the sum over the involution, $\sum_{\Gamma}f(\Gamma^\dagger)=\sum_\Gamma f(\Gamma)$, it results in
\begin{equation}
\label{dft2}
p(\sigma)=e^\sigma p(-\sigma).
\end{equation}
From (\ref{dft2}), we have the useful property \cite{Salazar2022c} for odd functions $u(-\sigma)=-u(\sigma$),
\begin{equation}
\label{oddproperty}
\langle u(\sigma) \rangle = \langle u(\sigma)\tanh(\sigma/2)\rangle.
\end{equation}
Now we decompose the mgf (\ref{mgf2}) in a sum of odd and even functions, $G(\alpha)=\langle \sinh(\alpha \sigma) \rangle + \langle \cosh(\alpha \sigma)\rangle$ and we use property (\ref{oddproperty}) for $u(\sigma)=\sinh(\alpha \sigma)$, which results in
\begin{eqnarray}
\label{mgf21}
G(\alpha)=\langle \sinh(\alpha \sigma)\tanh(\sigma/2)+\cosh(\alpha \sigma)\rangle.
\end{eqnarray}
Now we use the identity $\cosh(x+y)=\cosh(x)\cosh(y)+\sinh(x)\sinh(y)$ to obtain the compact form from (\ref{mgf21}),
\begin{equation}
\label{mgf3}
G(\alpha)= \big\langle \frac{\cosh[(\alpha + 1/2)\sigma]}{\cosh(\sigma/2)} \big\rangle.
\end{equation}
Finally, we define the functions $h(\sigma)=\sigma\tanh(\sigma/2)$ and the inverse $g(h(\sigma))=|\sigma|$. Inserting $|\sigma|=g(h(\sigma))$ in (\ref{mgf3}) leads to
\begin{equation}
\label{mgf4}
G(\alpha)= \Big\langle \frac{\cosh[(\alpha + 1/2)g(h(\sigma))]}{\cosh[(1/2)g(h(\sigma))]} \Big\rangle.
\end{equation}
The final step is to use Jensen's inequality in (\ref{mgf4}) in a strategy analogous to previous results \cite{Salazar2022c} as follows
\begin{equation}
\label{Jensens}
\langle f[g(h(\sigma))] \rangle \geq \langle f[g(\langle h(\sigma) \rangle)] = f[g(\langle \sigma \rangle)], 
\end{equation}
which is true if $w''(h):=d^2 w(h)/dh^2 > 0$, where $w(h):=f[g(h)]$ and $f(x)=\cosh[(\alpha+1/2)x]/\cosh(x/2)$ (see Appendix). Note that $\langle \sigma \rangle = \langle \sigma \tanh(\sigma/2)\rangle = \langle h(\sigma)\rangle$, from (\ref{oddproperty}) because $u(\Sigma)=\sigma$ is odd. Combining (\ref{mgf4}) and (\ref{Jensens}) and using $\langle \Sigma \rangle = \langle \sigma \rangle$, we obtain (\ref{main}).

{\bf \emph{Discussion-}} 
As in the case of previous results \cite{Timpanaro2019B,Salazar2022d}, it can be checked directly that the bound is saturated by the minimal distribution, $p(\Sigma)=[\delta(\Sigma+a)\exp(-a/2)+\delta(\Sigma-a)\exp(a/2)]/(2\cosh(a/2))$, where $a=g(\overline{\Sigma})$. 

An interesting aspect of (\ref{DFT}) is that $\Gamma$ could be replaced by the pair $\tilde{\Gamma}:=(d,\Gamma)$, where $\Gamma \in S$ is the original set and $d\in \{F,B\}$ denotes forward and backwards experiments, such that $m(\tilde{\Gamma})=m(d,\Gamma)=(d ^\dagger,\Gamma^\dagger)$, where $F^\dagger = B$ and $B^\dagger= F$, $P(\tilde{\Gamma}):=P_{d}(\Gamma)/2$. In this case, one has $\Sigma_d(\Gamma):= \Sigma(\tilde{\Gamma})=\ln P(d,\Gamma)/P(d^\dagger,\Gamma^\dagger)$, so that $\Sigma_F(\Gamma):=\ln P_F(\Gamma)/P_B(\Gamma)$, which is the usual definition of the detailed fluctuation theorem, as opposed to the strong DFT (\ref{DFT}). Thus, in the general framework presented in this paper, we do not make distinction between the detailed FT and the strong DFT.

We also point out that our main result (\ref{main}) could also be stated in terms of information theory as a bound for the Rényi relative entropy as follows. The definition of Rényi relative entropy for probabilities $P$ and $Q$ is
\begin{equation}
\label{relativeentropy}
S_{\alpha}(P|Q):=\frac{-1}{1-\alpha}\ln\Big( \sum_\Gamma P(\Gamma)^\alpha Q(\Gamma)^{1-\alpha}\Big).
\end{equation}
Now let ($S,P,m$) be defined as in the formalism, then $\overline{\Sigma}=D(P|P')$ is the Kullback-Leibler (KL) divergence, $D(P|Q):=\sum_i P_i \ln (P_i/Q_i)$, where $P'(\Gamma):=P(m(\Gamma))$. From (\ref{mgf}) and (\ref{relativeentropy}), we have $G(\alpha)=\exp[\alpha S_{\alpha+1}(P|P')]$. In terms of (\ref{relativeentropy}), the main result (\ref{main}) reads
\begin{equation}
\label{renyibound}
S_{\alpha}(P|P')\geq \frac{1}{\alpha-1}\ln \Big(\frac{\cosh[(\alpha-1/2)g(D(P|P'))]}{\cosh[(1/2)g(D(P|P'))]}\Big),
\end{equation}
for any set $S$, probability $P$ and involution $m$. Actually, for $\alpha\rightarrow 1$, the bound saturates as both sides of (\ref{renyibound}) converge to $D(P|P')$.  
One might wonder how a result (\ref{main}) that seemed dependent on the DFT is actually as general as (\ref{renyibound}). As it turns out, the involution property $m(m(\Gamma))=\Gamma$ constrains the pair of probabilities $(P,P')$ in (\ref{renyibound}), creating the same effect of the DFT. For instance, if $m$  is the identity, then $P=P'$ and, in thermodynamics, that would be equivalent to equilibrium (or detailed balance condition in the case of a Markov process). For a general involution $m$, probabilities $P$ and $P'$ will differ (nonequilibrium), as represented by the KL divergence $D(P|P')>0$ and the relative entropy $S_\alpha (P|P')>0$.

{\bf \emph{Gaussian case-}}
Now we compare the bound (\ref{main}) to some systems that satisfy the DFT. The Gaussian case \cite{Pigolotti2017,Chun2019} is given by the mgf
\begin{equation}
\label{gaussianbound}
G(\alpha)=\exp(\alpha(1+\alpha)\overline{\Sigma}),
\end{equation}
as the DFT fixes the variance $\langle \Sigma^2 - \overline{\Sigma}^2 \rangle =2 \overline{\Sigma}$ for the Gaussian distribution and it easily checks property (\ref{symmetry}). Interestingly, expression (\ref{gaussianbound}) appears as a lower bound for the mgf for steady states in stochastic thermodynamics \cite{Pietzonka2017a}, for the particular case where the current is the entropy production itself. The comparison between (\ref{gaussianbound}) and (\ref{main}) is depicted in Fig.~1, where $\ln G(\alpha)$ is quadratic in $\alpha$ for the Gaussian case (\ref{gaussianbound}), but it shows higher order corrections in our bound (\ref{main}). 

\begin{figure}[htp]
\includegraphics[width=3.3 in]{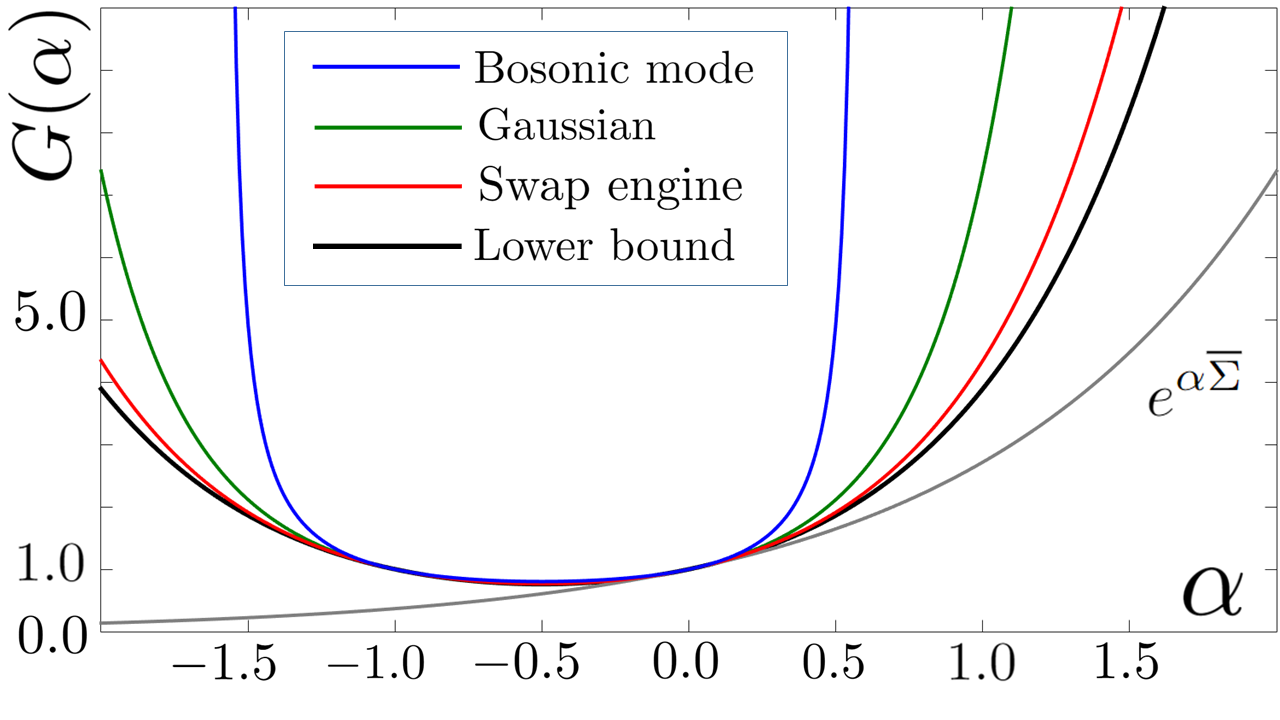}
\caption{(Color online) Moment generating function (MGF) $G(\alpha)$ as a function of $\alpha$ for $\overline{\Sigma}=1$ for the lower bound (black), swap engine (red), Gaussian (green) and bosonic mode (blue). We also show the exponential $\exp(\alpha \overline{\Sigma})$ in gray. Because of the DFT, note that all MGFs are symmetric around $\alpha=-0.5$ (\ref{symmetry}) and the integral FT is also verified $G(-1)=G(0)=1$. The MGF of the bosonic mode is not defined for the entire domain (see applications). The Gaussian case, which is a quadratic form for $\ln G(\alpha)$, and the swap engine departs from the bound for large $|\alpha|$.}
\label{fig1}
\end{figure}

{\bf \emph{Bosonic mode case-}} For another comparison, we take a free bosonic mode with Hamiltonian $H=\hbar\omega (a^\dagger a+1/2)$ weakly coupled to a thermal reservoir with a density matrix satisfying a Lindblad's equation as developed in previous results \cite{Santos2017a,Salazar2019,Denzler2019},
\begin{equation}
\label{Lind}
\partial_t \rho = \frac{-i}{\hbar}[H,\rho] + D_i(\rho),
\end{equation}
for the dissipator given by
\begin{equation}
D_i(\rho)=\gamma(\overline{n}_i+1)[a\rho a^\dagger - \frac{1}{2}\{a^\dagger a, \rho\}]+\gamma \overline{n}_i[a^\dagger \rho a -\frac{1}{2}\{a a^\dagger , \rho\}],
\end{equation}
    where $\gamma$ is a constant and $\overline{n}_i=[\exp(\hbar \omega/k_BT_i)-1]^{-1}$ is the bosonic thermal occupation number and $\beta_i=1/(k_B T_i)$. We denote the solution of (\ref{Lind}) as $\rho_t:= \Phi_t (\rho_0)$. A two point measurement scheme is performed: first, the system is prepared in thermal equilibrium (temperature $T_1$) at time $t=0$, when the first energy measurement takes place yielding $E_1$. Then, the system is placed in thermal contact with a second reservoir (temperature $T_2$), when another energy measurement is performed (time $t>0$) yielding $E_2$. The dynamics of the system between the two measurements is given by (\ref{Lind}) with temperature $T_2$. As explored in previous papers, repeating the experiment multiple times results in a distribution for $\Delta E = E_2 - E_1$, where $\Sigma := -(\beta_2-\beta_1)\Delta E$ \cite{Campisi2015,Timpanaro2019B,Sinitsyn2011} is the entropy production. In this case, the distribution is given by $p(\Delta E=\hbar \omega m)=\sum_{n=0}^{\infty}\langle n+m | \Phi_t \big(|n\rangle \langle n |\big)|n+m\rangle p_n$, where $p_n={e^{-\beta_1 E_n}}/Z(\beta_1)$ and $Z(\beta_1)=\tr (e^{-\beta_1 H})$, resulting in the closed form \cite{Salazar2019,Denzler2019} for the entropy production
\begin{equation}
\label{PsigmaHO}
    p(\Sigma)=\frac{1}{A(0,\lambda)} \exp(\frac{\Sigma}{2}-\lambda \frac{|\Sigma|}{2}),
\end{equation}
with support $s=\{\pm \Delta \beta \hbar \omega m\}=\{\pm\varepsilon m\}$, $m=0,1,2,..$, and normalization constant $A(0,\lambda)$, where $A(\alpha,\lambda):=1+\sum_{m=1}^\infty [\exp((\alpha+m/2)\epsilon)-\exp(-\alpha-m/2)\epsilon)]\exp(-\lambda \epsilon m/2))$, for $\lambda > |2\alpha+1|$. Upon inspection, note that (\ref{PsigmaHO}) checks (\ref{DFT}). The MGF (\ref{mgf}) for (\ref{PsigmaHO}) is given by
\begin{equation}
    \label{mgfHO}
 G(\alpha)=\frac{A(\alpha,\lambda)}{A(0,\lambda)},
\end{equation}
for $\lambda > |2\alpha+1|$. In Fig.1, we plot the MGF (\ref{mgfHO}) or several values of $\alpha$, with $\varepsilon=1$ and $\lambda=\lambda^*$ such that $\langle \Sigma\rangle=1$, compared to the lower bound (\ref{main}) for the same $\overline{\Sigma}=1$. This example shows a case where the domain of $\alpha$ is limited, $\lambda^* > |2\alpha+1|$, but the bound is still satisfied.

{\bf \emph{Swap engine -}} Consider a pair of qubits with energy gaps $\epsilon\in\{\epsilon_A,\epsilon_B\}$. They are prepared in thermal equilibrium, $p(\pm)=\exp(\pm\beta \epsilon)/(\exp(-\beta\epsilon)+\exp(+\beta\epsilon))$, for $\beta\in\{\beta_1,\beta_2\}$, with reservoirs at temperatures $T_1 , T_2$. A TPM is performed before and after a swap operation \cite{Campisi2015}, $|xy\rangle \rightarrow |yx\rangle$, for $x,y \in \{-,+\}$. The entropy production is given \cite{Campisi2015,Timpanaro2019B} by $\Sigma = \beta_1 \Delta E_A + \beta_2 \Delta E_B$,
where $\Delta E_A = E_A^f-E_A^i$, $\Delta E_B=E_B^f-E_B^i$ are the variations of energy measurements before and after the swap. In this TPM, the outcomes are $\Sigma \in s=\{0,\pm 2a\}$ for $2a=2(\beta_2\epsilon_B-\beta_1\epsilon_A)$. The distribution of $\Sigma$ is given by 
\begin{equation}
\label{pdfswap}
p(\Sigma)=(1/Z_0)[\delta(\Sigma)+\delta(\Sigma+2a)e^{-a}+\delta(\Sigma-2a)e^{a}],
\end{equation}
for $Z_0=1+\exp(a)+\exp(-a)$, which satisfies the DFT (\ref{DFT}). The MGF of (\ref{pdfswap}) is given by
\begin{equation}
\label{swapmgf}
G(\alpha) = \frac{1+2\cosh(a(2\alpha+1))}{1+2\cosh(a)},
\end{equation}
where $a$ defines $\overline{\Sigma}$ uniquely from $\overline{\Sigma}=2a(\exp(a)-\exp(-a))/(1+\exp(a)+\exp(-a))$. Comparison between (\ref{swapmgf}) and (\ref{main}) is also depicted in Fig.1 for $\overline{\Sigma}=1$.

{\bf \emph{Conclusions -}}
We explored the DFT (\ref{DFT}) as a definition of a random variable $\Sigma$ and proved the impact it has on the MGF (\ref{mgf}). The result is that the MGF of $\Sigma$ is lower bounded as a function of the mean $\overline{\Sigma}$ and the parameter $\alpha$. This lower bound improves on the simple exponential bound $\exp(\alpha \overline{\Sigma})$ because of the special definition of $\Sigma$ in terms of the DFT. Although the structure (\ref{DFT}) appears as a definition of entropy production in stochastic thermodynamics, it is also the case of general entropy related quantities (that might include boundary and information terms) in quantum thermodynamics even beyond two point measurement schemes \cite{Micadei2019}. We also wrote the main result in terms of information theory, as a lower bound for the Rényi relative entropy between distributions $P$ and $P'$, where the pair is constrained by an involution, $P'(\Gamma)=P(m(\Gamma))$.

{\bf \emph{Appendix-}}
For the proof of (\ref{Jensens}), consider the  notation $w':=dw/dh$ and $\dot{w}:=dw/d\sigma$.
We have $w'=\dot{w}\sigma'$ and
\begin{eqnarray}
\label{app1}
    w''=\frac{d}{dh}(\dot{w}\sigma')=\ddot{w}\sigma'^2+\dot{w}\sigma'',
\end{eqnarray}
where $\sigma'=d\sigma/dh=1/\dot{h}$ and $\sigma''=\sigma'(d/d\sigma)(1/\dot{h})=-\ddot{h}\sigma'/\dot{h}^2=-\ddot{h}/\dot{h}^3$. Replacing $\sigma'$ and $\sigma''$ in (\ref{app1}) yields 
\begin{eqnarray}
\label{app2}
    w''=\frac{1}{{\dot{h}^2}}\big(\ddot{w}-\dot{w}\frac{\ddot{h}}{\dot{h}}\big).
\end{eqnarray}
Finally, using $h(\sigma)=\sigma\tanh(\sigma/2)$ and $w(h(\sigma))=\cosh((\alpha+1/2)\sigma))/\cosh(\sigma/2)$ explicitly to calculate $\ddot{w}, \dot{w}$, $\dot{h}$ and $\ddot{h}$, one obtains $F(\sigma):=\dot{h}^2 w''$ from (\ref{app2})
\begin{eqnarray}
\label{app3}
    F(\sigma)=\frac{(\alpha^2+\alpha^2 j(\sigma)+\alpha)f(\sigma)-(2\alpha^2+\alpha) j(\alpha \sigma)}{1+j(\sigma)},
\end{eqnarray}
where $f(x):=\cosh((\alpha+1/2)x)/\cosh(x/2)$ and $j(x)=\sinh(x)/x$. Now we use $j(\sigma)=\sinh(\sigma)/\sigma \geq 1$, which in combination with $\alpha^2 \geq 0$, $f(x)\geq 0$ and $j(x) \geq 0$ results in
\begin{equation}
\label{app4}
F(\sigma) \geq \frac{(2\alpha+1)}{1+j(\sigma)}\big(\alpha f(\sigma)-\alpha j(\alpha \sigma)\big).
\end{equation}
We are interested in $2\alpha + 1 \geq 0$, as the region $\alpha < -1/2$ is obtained with the reflection $G(\alpha)=G(-\alpha-1)$. In the case $\alpha > -1/2$, we first consider $\alpha \geq 0$, for which (\ref{app4}) is rewritten as
\begin{equation}
\label{app5}
F(\sigma) \geq \frac{(2\alpha+1)}{1+j(\sigma)}\big(\alpha \cosh(\alpha \sigma)-\frac{\sinh(\alpha \sigma)}{\sigma} + \alpha \sinh(\alpha \sigma)\tanh(\sigma/2)\big).
\end{equation}
Now we use $\cosh(\alpha x) \geq \sinh(\alpha x)/\alpha x$ and $\alpha\geq 0$ to get from (\ref{app5})
\begin{equation}
\label{app6}
F(\sigma) \geq \frac{(2\alpha+1)}{1+j(\sigma)}\big(\alpha \sinh(\alpha \sigma)\tanh(\sigma/2)\big) \geq 0.
\end{equation}
Now for the case $\alpha \in (-1/2,0)$, we rewrite (\ref{app4}) as 
\begin{equation}
\label{app7}
F(\sigma) \geq \frac{(-\alpha)(2\alpha+1)}{1+j(\sigma)}\frac{\big(\cosh(\sigma/2)-\cosh(|\alpha+1/2|\sigma)\big)}{\cosh(\sigma/2)} \geq 0,
\end{equation}
since $-\alpha \geq 0$ and $\cosh(\sigma/2)- \cosh(|\alpha+1/2|\sigma)\geq0$ for $0<|\alpha+1/2|<1/2$. Combining (\ref{app6}) and (\ref{app7}), we have $F(\sigma)=\dot{h}^2 w''\geq 0$ for $\alpha \geq -1/2$, which results in $w''\geq 0$ for $\alpha\geq -1/2$. It allows the lower bound to be written for $\alpha\geq -1/2$ as
\begin{equation}
\label{mainapp}
G(\alpha) \geq \frac{\cosh((\alpha+1/2)g(\langle \sigma \rangle))}{\cosh(g(\langle \sigma \rangle)/2)}:=B(\alpha,\langle \sigma \rangle).
\end{equation}
Now for $\alpha<-1/2$ we could
use the symmetry (\ref{symmetry}),
\begin{equation}
G(\alpha)=G(-1-\alpha)\geq B(-1-\alpha,\langle \sigma \rangle)=B(\alpha,\langle \sigma \rangle),
\end{equation}
where we used the property $B(\alpha,x)=B(-1-\alpha,x)$ and (\ref{mainapp}) in the inequality above as $\alpha<-1/2 \rightarrow -1-\alpha>-1/2$. That completes the proof for $\alpha \in \mathbb{R}$.

\bibliography{lib6}
\end{document}